

\font\titlefont = cmr10 scaled\magstep 4
\font\sectionfont = cmr10

\font\teenyfont = cmr5

\magnification = 1200

\global\baselineskip = 1.2\baselineskip
\global\parskip = 4pt plus 0.3pt
\global\nulldelimiterspace = 0pt

\predisplaypenalty 1000


\def\endignore{}
\def\ignore #1\endignore{}

\newcount\dflag
\dflag = 0


\def\monthname{\ifcase\month
\or Jan \or Feb \or Mar \or Apr \or May \or June%
\or July \or Aug \or Sept \or Oct \or Nov \or Dec
\fi}



\def\endid{}
\def\id#1\endid{\number\day\ \monthname \number\year
\hfill #1}

\def\endtitle{}
\def\title#1\endtitle{\vskip.3in\titlefont
\global\baselineskip = 2\baselineskip
#1\vskip.4in
\baselineskip = 0.5\baselineskip\rm}


\def\endauthors{}
\def\authors#1\endauthors{#1}

\def\endabstract{}
\def\abstract#1\endabstract{\vskip .3in%
\centerline{\sectionfont\bf Abstract}%
\vskip .1in%
\noindent#1%
\ifnum\dflag = 0
\footline = {\hfil}\pageno = 0
\vfill\eject
\fi}


\newcount\nsection
\newcount\nsubsection

\def\section#1{\global\advance\nsection by 1
\global\nsubsection = 0
\bigskip\noindent\sectionfont \bf \number\nsection.\ #1
\nobreak\medskip\rm\nobreak}

\def\subsection#1{\global\advance\nsubsection by 1
\bigskip\noindent\sectionfont \it \number\nsection.\number\nsubsection.\ #1%
\nobreak\medskip\rm\nobreak}

\def\appendix#1#2{\bigskip\noindent%
\sectionfont \bf Appendix #1.\ #2
\nobreak\medskip\rm\nobreak}


\newcount\nref
\global\nref = 1

\def\ref#1#2{\xdef #1{[\number\nref]}
\ifnum\nref = 1\global\xdef\therefs{\noindent[\number\nref] #2\ }
\else
\global\xdef\oldrefs{\therefs}
\global\xdef\therefs{\oldrefs\vskip.1in\noindent[\number\nref] #2\ }%
\fi%
\global\advance\nref by 1
}

\def\listrefs{\vfill\eject\section{References}\therefs}


\newcount\nfig
\global\nfig = 1

\def\fg#1\efig{\vskip .5in\noindent Fig.\ \number\nfig:\ #1%
\global\advance\nfig by 1}


\newcount\cflag
\newcount\nequation
\global\nequation = 1
\def\eqlabel{(1)}

\def\nexteqno{\ifnum\cflag = 0
\global\advance\nequation by 1
\fi
\global\cflag = 0
\xdef\eqlabel{(\number\nequation)}}

\def\lasteqno{\global\advance\nequation by -1
\xdef\eqlabel{(\number\nequation)}}

\def\label#1{\xdef #1{(\number\nequation)}
\ifnum\dflag = 1
{\escapechar = -1
\xdef\draftname{\teenyfont\string#1}}
\fi}

\def\clabel#1#2{\xdef\eqlabel{(\number\nequation #2)}
\global\cflag = 1
\xdef #1{\eqlabel}
\ifnum\dflag = 1
{\escapechar = -1
\xdef\draftname{\string#1}}
\fi}

\def\cclabel#1#2{\xdef\eqlabel{#2)}
\global\cflag = 1
\xdef #1{\eqlabel}
\ifnum\dflag = 1
{\escapechar = -1
\xdef\draftname{\string#1}}
\fi}


\def\eeq{}

\def\eqnn #1\eeq{$$ #1 $$}

\def\eq #1\eeq{\xdef\draftname{\ }
$$ #1
\eqno{\eqlabel \rlap{\ \draftname}} $$
\nexteqno}



\def\eol{& \eqlabel \rlap{\ \draftname} \crcr
\nexteqno
\xdef\draftname{\ }}

\def\eeol{& \eqlabel \rlap{\ \draftname}
\nexteqno
\xdef\draftname{\ }}

\def\eolnn{\cr
\global\cflag = 0
\xdef\draftname{\ }}


\def\eqa #1\eeq{\xdef\draftname{\ }
$$ \eqalignno{ #1 } $$
\global\cflag = 0}


\def\ie{{\it i.e.\/}}
\def\eg{{\it e.g.\/}}

\def\myinstitution{
    \centerline{\it Theoretical Physics Group}
    \centerline{\it Lawrence Berkeley Laboratory}
    \centerline{\it 1 Cyclotron Road}
    \centerline{\it Berkeley, California 94720}
}


\def\jref#1#2#3#4{{\it #1} {\bf #2}, #3 (#4)}

\def\MPLA#1#2#3{\jref{Mod.\ Phys.\ Lett.}{A#1}{#2}{#3}}

\def\NPB#1#2#3{\jref{Nucl.\ Phys.}{B#1}{#2}{#3}}

\def\PLB#1#2#3{\jref{Phys.\ Lett.}{#1B}{#2}{#3}}
\def\PR#1#2#3{\jref{Phys.\ Rep.}{#1}{#2}{#3}}
\def\PRD#1#2#3{\jref{Phys.\ Rev.}{D#1}{#2}{#3}}

\def\PRL#1#2#3{\jref{Phys.\ Rev.\ Lett.}{#1}{#2}{#3}}
\def\PRV#1#2#3{\jref{Phys.\ Rev.}{#1}{#2}{#3}}
\def\PTP#1#2#3{\jref{Prog.\ Theor.\ Phys.}{#1}{#2}{#3}}


\def\goto{\mathop{\rightarrow}}
\def\gotoo{\mathop{\longrightarrow}}


\def\myint{\int\mkern-5mu}
\def\frac#1#2{{{#1} \over {#2}}\,}  
\def\sfrac#1#2{{\textstyle\frac{#1}{#2}}}  


\def\Dsl{\hbox{/\kern-.6000em\rm D}} 



\def\rsub#1{\mathstrut_{\rm #1}}

\def\twi{\widetilde}
\def\mybar#1{\kern 0.8pt\overline{\kern -0.8pt#1\kern -0.8pt}\kern 0.8pt}
\def\sla#1{\raise.15ex\hbox{$/$}\kern-.57em #1}
\def\Sla#1{\kern.15em\raise.15ex\hbox{$/$}\kern-.72em #1}

\def\roughly#1{\mathrel{\raise.3ex\hbox{$#1$\kern-.75em%
    \lower1ex\hbox{$\sim$}}}}
\def\lsim{\roughly<}

\def\scr#1{{\cal #1}}






\def\tr{\mathop{\rm tr}}



\def\vev#1{\langle 0 | #1 | 0 \rangle}
\def\avg#1{\langle #1 \rangle}



\def\rhs{right-hand side}

\def\hc{{\rm h.c.}}


\def\GeV{{\rm \ GeV}}
\def\TeV{{\rm \ TeV}}


\id
LBL-32089
\endid

\title
\centerline{Electroweak Symmetry Breaking via a}
\centerline{Technicolor Nambu--Jona-Lasinio Model}
\endtitle

\authors
\centerline{Markus A. Luty}
\vskip .1in
\myinstitution
\endauthors

\abstract
We consider a theory of gauge fields and fermions which we argue gives rise
to dynamics similar to that of the Nambu--Jona-Lasinio (NJL) model when a gauge
coupling constant is appropriately fine-tuned.
We discuss the application of this model to dynamical electroweak symmetry
breaking by a top-quark condensate.
In this model, custodial symmetry is violated solely by perturbatively weak
interactions, and the top--bottom mass splitting is due to the enhanced
sensitivity to custodial symmetry violation near the critical point.
We also consider models in which electroweak symmetry is broken by new
strongly-interacting fermions with NJL-like dynamics.
We argue that these models require additional fine-tuning in
order to keep corrections to the electroweak $\rho$-parameter acceptably
small.
\endabstract

\def\GTC{$SU(N)_{TC}$}
\def\GC{$SU(N)_C$}
\def\GI{$\left[ SU(K) \times SU(K) \right]_I$}
\def\GII{$SU(K)_{I'}$}
\def\GW{$SU(2)_W \times U(1)_Y$}

\def\Gc{G\rsub{crit}}

\def\SD{Schwinger--Dyson}

\ref\NJL{G.\ Jona-Lasinio and Y.\ Nambu, \PRV{122}{345}{1961}.}

\ref\topmode{Y.\ Nambu, EFI-88-39 (1988), EFI-88-62 (1988), and
EFI-89-08 (1989) (unpublished);
V.\ A.\ Miransky, M.\ Tanabashi, and K.\ Yamawaki, \MPLA{4}{1043}{1989};
\PLB{221}{177}{1989};
W.\ Marciano, \PRL{62}{2793}{1989}.}

\ref\BHL{W.\ A.\ Bardeen, C.\ T.\ Hill, and M.\ Lindner,
\PRD{41}{1647}{1990}.}

\ref\topvar{For example, see M.\ A.\ Luty, \PRD{41}{2893}{1990};
M.\ Suzuki, \PRD{41}{3457}{1990};
C.\ T.\ Hill, M.\ A.\ Luty, and E.\ A.\ Paschos, \PRD{43}{3011}{1991};
R.\ N.\ Mohapatra and K.\ S.\ Babu, \PRL{66}{556}{1991};
P.\ H.\ Frampton, O.\ Yasuda, \PRD{44}{3709}{1991};
M.\ Suzuki, \PRD{44}{3628}{1991};
W.\ A.\ Bardeen, M.\ Carena, T.\ E.\ Clark, K.\ Sasaki, and C.\ Wagner,
\NPB{369}{33}{1992}.}

\ref\renorm{C.\ T.\ Hill, \PLB{266}{419}{1991};
D.\ E.\ Clague and G.\ G.\ Ross, \NPB{364}{43}{1991};
M.\ Lindner and D.\ Ross, \NPB{370}{30}{1992};
S.\ P.\ Martin, Florida Univ.\ preprint UFIFT-HEP-91-24;
R.\ B\"onish, \PLB{268}{394}{1991}.}

\ref\Suzuki{M.\ Suzuki, \MPLA{5}{1205}{1990}}

\ref\Hasens{A.\ Hasenfratz, P.\ Hasenfratz, K.\ Jansen, J.\ Kuti, and
Y.\ Shen, \NPB{365}{79}{1991}.}

\ref\Chrisfix{C.\ T.\ Hill, \PRD{24}{691}{1981};
C.\ T.\ Hill, C.\ Leung, and S.\ Rao, \NPB{262}{517}{1985}.}

\ref\align{S.\ Chadha and M.\ E.\ Peskin, \NPB{185}{61}{1981}, and references
therin.}

\ref\gaprev{For reviews with various points of view, see
T.\ Appelquist, Yale preprint YCTP-P23-91, to be published in the proceedings
of
the Fourth Mexican Summer School of Particles and Fields;
M.\ Peskin in {\it Recent Advances in Field Theory and Statistical Mechanics,}
edited by R.\ Stora and J.\ B.\ Zuber (Elsevier, Amsterdam, 1982);
A.\ G.\ Cohen and H.\ Georgi, \NPB{314}{7}{1989}.}

\ref\JJ{R.\ Jackiw and K.\ Johnson, \PRD{8}{2386}{1973};
For a recent discussion, see T.\ Appelquist, M.\ B.\ Einhorn, T.\ Takeuchi,
and L.\ C.\ R.\ Wijewardhana, \PRD{41}{3192}{1990}.}

\ref\PZ{E.\ A.\ Paschos and V.\ I.\ Zakharov, \PLB{272}{105}{1991}.}

\ref\TCNJL{R.\ S.\ Chivukula, A.\ G.\ Cohen, and K.\ Lane,
\NPB{343}{554}{1990}.}

\ref\crit{T.\ Appelquist and O.\ Shapira, \PLB{249}{83}{1990};
T.\ Appelquist, J.\ Terning, and L.\ C.\ R.\ Wijewardhana, \PRD{44}{}{1991}.}

\ref\ttcrit{M.\ Bando, T.\ Kugo, and K.\ Suehiro, \PTP{85}{1299}{1991}.}

\ref\axionrev{For reviews, see R.\ Peccei in {\it CP Violation}, edited by
C.\ Jarlskog (World Scientific, Singapore, 1989);
J.\ Kim, \PR{150}{1}{1987}.}

\ref\PDG{Particle Data Group, private communication.}

\section{Introduction}

In the last few years, there has been a revived interest in the
Nambu--Jona-Lasinio (NJL) model \NJL\ as an effective description of
electroweak symmetry breaking via a top-quark condensate \topmode\BHL.
Numerous variations on this basic theme have also been explored \topvar.
More recently, several groups have constructed renormalizable models which
are supposed to give rise to top-quark condensates and NJL-like
dynamics. \renorm.
In this paper, we will consider a model of asymptotically free gauge fields
and fermions which we argue has dynamics similar to that of the NJL model.
In this model, all symmetry breaking is driven by strongly-coupled $SU(N)$
gauge groups with fermions in the fundamental representation.
We can therefore use our understanding of chiral symmetry breaking in
QCD-like theories to analyze the model.
The only new dynamical assumption is that the chiral symmetry
breaking phase transition is second order.
We will consider models in which the electroweak symmetry is broken by
top-quark condensates, as well as models in which electroweak symmetry is
broken by condensates of new fermions.

The plan of this paper is as follows:
In section 2, we will review some basic results concerning the NJL model, and
discuss the sense in which the NJL model can be viewed as a low-energy
effective lagrangian.
In section 3, we will present a model of asymptotically-free gauge fields and
fermions and give qualitative arguments that this model gives rise to dynamics
similar to the NJL model.
In section 4, we flesh out the dynamical picture of the previous section by
analyzing the model using the \SD\ equations in ladder approximation.
In section 5, we apply the model to electroweak symmetry breaking, and section
6
contains our conclusions.

\section{The NJL model}

We begin by treating the NJL model as a toy model with no reference to
electroweak physics.
The NJL model we will consider has $F$ ``flavors'' and $N$ ``technicolors.''
The lagrangian is
\eq
\label\NJLL
\scr L\rsub{NJL} = \mybar\psi i\sla\partial \psi
+ \frac GN \; (\mybar\psi_{Lja} \psi_{Rka}) (\mybar\psi_{Rkb} \psi_{Ljb}),
\eeq
where $j,k = 1, \ldots, F$ are flavor indices and $a,b = 1, \ldots, N$ are
technicolor indices.
This theory has a global chiral symmetry
\eq
\label\chisymm
SU(N)_{TC} \times U(F)_{\psi L} \times U(F)_{\psi R},
\eeq
in an obvious notation.
The theory is nonrenormalizable, and must therefore be interpreted as an
effective theory equipped with an energy cutoff $\Lambda$.
In order to describe physics at scales above $\Lambda$, we must construct
a more fundamental theory which reduces to the NJL model at low
energies.

In the limit $N \goto \infty$ (with $F$ held finite), this model can be
solved exactly.
The qualitative features are well known:
if $G$ is less than a critical value $\Gc$, then the chiral symmetry of
eq.\ \chisymm\ is unbroken, and the effective theory below the scale
$\Lambda$ simply describes massless fermions interacting through the contact
interaction of the form eq.\ \NJLL.
However, if $G > \Gc$, then the chiral symmetry is partially broken:
the condensate
\eq
\label\cond
\avg{\mybar\psi_{Lja} \psi_{Rja}} \ne 0
\eeq
results in the symmetry-breaking pattern
\eq
\label\NJLpatt
U(F)_{\psi L} \times U(F)_{\psi R} \gotoo U(F)_{\psi L + \psi R},
\eeq
where $U(F)_{\psi L + \psi R}$ is the vector-like subgroup of
$U(F)_{\psi L} \times U(F)_{\psi R}$.

If $G$ is tuned to be very close to $\Gc$,
\eq
G = \Gc \left[ 1 + O(G m^2) \right]
\ \ \ {\rm with}\ \ \ m^2 \ll G^{-1},
\eeq
then the effective theory near the scale $m$ is precisely a
$U(F) \times U(F)$ linear sigma model coupled to the fermions $\psi$.
The effective lagrangian at the scale $m$ is
\eqa
\scr L_{\rm eff} & = \mybar\psi i\sla\partial \psi
+ y \left( \mybar\psi_{Lj} \Phi_{jk} \psi_{Rk} + \hc \right) \eolnn
\label\efflin
& \quad + \tr \partial^\mu \Phi^\dagger \partial_\mu \Phi
- \mu_\Phi^2 \tr \Phi^\dagger \Phi
- \lambda_1 \left( \tr \Phi^\dagger \Phi \right)^2
- \lambda_2 \tr \Phi^\dagger \Phi \Phi^\dagger \Phi
+ \cdots, \eeol
\eeq
where the ellipses indicate terms containing higher-dimension operators.
If $G < \Gc$, then $\mu_\Phi^2 > 0$ and the theory is in the unbroken phase.
The fermions are then massless, and there are $2F^2$ physical scalars with
equal masses of order $m$.
If $G > \Gc$, then $\mu_\Phi^2 < 0$ and the theory is in the broken phase.
The fermions then have masses of order $m$, and there are $F^2$ physical
scalars with masses of order $m$ and $F^2$ massless Nambu--Goldstone bosons.
Because the underlying theory eq.\ \NJLL\ has only one coupling constant $G$,
there will be non-trivial relations among the parameters of the effective
low-energy lagrangian eq.\ \efflin.

An important feature of the NJL model in the large-$N$ limit is that the
low-energy dynamics is continuous as a function of $G$ near the critical
point.
An analogous assumption in the context of our renormalizable model will play
an important role in what follows.

Before presenting the renormalizable model, we discuss the sense in which the
NJL model can arise as an effective theory from a more fundamental theory.
When we write an effective field theory which matches onto a more fundamental
theory at a scale $\Lambda$, we must include all possible interaction terms
among the low-energy fields consistent with the symmetries of the underlying
theory.
The coefficients of these operators are determined by matching the predictions
of the low-energy theory with that of the underlying theory.
There will be an infinite number of operators in the low-energy effective
lagrangian, but only a few of these, the so-called ``relevant'' operators,
will be important for describing the physics at energies $\mu \ll \Lambda$.
The effects of the remaining ``irrelevant'' operators at low energies will be
suppressed by powers of $\mu / \Lambda$.
This is the reason for the power of the effective-lagrangian approach.

For weakly-coupled theories, dimensional analysis is sufficient to classify
operators as relevant or irrelevant:
operators with engineering dimension 4 or less are relevant, and all others
are irrelevant.
As the NJL model illustrates, the classification becomes non-trivial when
couplings become strong.
We therefore ask if there are any other operators which are relevant if they
are added to the NJL lagrangian eq.\ \NJLL\ when $G$ is near its critical
value.

A clue to the answer to this question can be obtained from the fact that the
low-energy limit of the fine-tuned NJL model is a linear sigma model.
This equivalence can be made explicit by rewriting the NJL interaction in
terms of an auxilliary scalar field $\Phi$:
\eq
\scr L'\rsub{NJL} = \mybar\psi i \sla\partial \psi
- G^{-1} \Phi^*_{jk} \Phi_{jk}
+ \left( \mybar\psi_{Lja} \Phi_{jk} \psi_{Rka} + \hc \right).
\eeq
This is the effective lagrangian at the scale $\Lambda$, in which $\Phi$ has
no kinetic or self-interaction terms.
$\Phi$ can be explicitly integrated out to recover the NJL lagrangian
eq.\ \NJLL.
At scales below $\Lambda$, $\Phi$ acquires both kinetic and self-interaction
terms as in eq.\ \efflin, and can serve as an interpolating field for the
physical scalar states.

In this formulation of the NJL model, it is clear that {\it all} of the
the dimension-four operators appearing in eq.\ \efflin\ are relevant.
If these are added to the effective lagrangian at the scale $\Lambda$ and the
field $\Phi$ is integrated out, the result will be a non-local effective
action containing only the $\psi$ fields.
If we expand this action in powers of derivatives of $\psi$ fields, it seems
clear that there should be a small number of higher-dimension operators that
become relevant in the fine-tuned limit, and whose effects at low energies
precisely reproduce the effects of the operators of eq.\ \efflin.

One such operator was considered in ref.\ \Suzuki.
A more complete analysis was performed in ref.\ \Hasens, which explicitly
identified operators that, when added to the effective lagrangian at the scale
$\Lambda$, reproduced the entire parameter space of the linear sigma model.
It may therefore seem that the predictions of refs.\ \topmode\BHL\topvar\
based on the NJL model are vacuous, but this is not so.
The reason is that as long as $\Lambda$ is many orders of magnitude above
the weak scale, and the linear sigma model has a large Yukawa coupling at
scales just below $\Lambda$, the logarithmic evolution of the parameters of
the effective lagrangian drives the parameters of the theory to an approximate
infrared fixed point at the weak scale \Chrisfix.
(This is true for all models which reduce to the standard model with one or
two Higgs doublets in the low-energy limit.)
The role of the NJL-like dynamics in these models is therefore to explain
why the top-quark coupling is singled out to be strong enough to lie in the
basin of attraction of the approximate infrared fixed point.

The situation is clearly quite different if the compositeness scale is taken
to be near the weak scale.
In this case, the couplings do not run over a sufficient range to be
affected by the approximate infrared fixed point, and all the relevant
operators are {\it \'a priori} equally important.
The construction of explicit renormalizable models is clearly especially
interesting in this case.

If we consider some fundamental theory which gives rise to the NJL interaction
eq.\ \NJLL, it is clear that {\it all} of the relevant operators discussed
above will be generated.
Therefore, in attempting to construct a renormalizable model which captures
the physics of the NJL model, strictly speaking we can only expect to find a
model with a fermion bilinear order parameter whose low-energy limit is a
linear sigma model.
Nonetheless, in the model we construct, we will find that the dynamics is very
similar to the NJL model eq.\ \NJLL.

\section{A Renormalizable Model}

The NJL model described in the previous section bears a superficial
resemblace to a model in which the four-fermion coupling in eq.\ \NJLL\ is
replaced by \GTC\ gauge interactions.
In the \GTC\ gauge model, the gauge coupling becomes strong at some scale
$\Lambda_{TC}$, and a condensate of the form eq.\ \cond\ forms, giving rise to
$F^2 - 1$ NGB's.
(Recall that the NJL model of eq.\ NJLL\ has $F^2$ NGB's.)
We can try to press the analogy further by using Fierz indentities to
rewrite the NJL interaction as
\eq
\label\Fint
(\mybar\psi_{Lja} \psi_{Rka}) (\mybar\psi_{Rkb} \psi_{Ljb})
= (\mybar\psi_{Lj} \gamma^\mu T_A \psi_{Lj})
(\mybar\psi_{Rk} \gamma^\mu T_A \psi_{Rk}) + O(1 / N),
\eeq
where the $T_A$ are \GTC\ generators normalized so that
$\tr T_A T_B = \delta_{AB} / 2$.
Eq.\ \Fint\ is the operator corresponding to the most attractive channel
for one gauge boson exchange in the \GTC\ theory.
However, the two models are clearly qualitatively different:
The NJL model contains propagating fermions at low energies, while the
fermions are confined in the \GTC\ gauge model.
Also, the NJL model can be fine-tuned to make the scalar and fermion masses
small compared to the cutoff scale $\Lambda$, while the \GTC\ gauge model has
no adjustable parameter; all dimensionful quantities in the \GTC\ model are
of order $\Lambda_{TC}$ raised to the appropriate power.

We can, however, contruct a class of models whose low-energy dynamics
interpolates continuously between that of the NJL model and the \GTC\ gauge
model.
The models we will construct consist entirely of gauge fields coupled to
fermions, so the theories are well-behaved at high energies.
Chiral symmetry breaking in these models is triggered by asymptotically-free
gauge couplings, and the fine-tuning in the NJL limit can be understood in
terms of the fine-tuning of gauge couplings (defined at a suitable subtraction
point) in the underlying gauge theory.

The basic idea underlying the model is very simple.
We assume that there is another ``interloping'' sector which breaks the
\GTC\ group {\it completely}, so that all the \GTC\ gauge bosons are massive.
If this symmetry breaking occurs at a scale where the \GTC\ coupling is weak,
the effective theory below this scale will consist of massless fermions
interacting through contact interactions such as those of eq.\ \Fint, and
$\avg{\mybar\psi\psi} = 0$.
On the other hand, if the \GTC\ coupling gets strong at a scale higher than
the scale at which the interloping sector becomes strong, then the \GTC\
interactions will break the $\psi$ chiral symmetry and
$\avg{\mybar\psi\psi} \sim \Lambda_{TC}^3$.
In this case, the low-energy effective theory contains only the
Nambu--Goldstone bosons resulting from the chiral symmetry breaking.
If we assume that the transition between the two limits just described is
smooth, then we can obtain a model in which
$\avg{\mybar\psi\psi} \ll \Lambda_{TC}^3$ by tuning the model between the two
limits described above.
We expect the dynamics of such a model to be similar to the fine-tuned NJL
model discussed in the previous section.

Actually, since we are ultimately interested in obtaining a condensate for
quarks carrying only color indices, we will consider models where the
technicolor group is broken in the pattern
\eq
SU(N)_{TC} \times SU(N)_C \gotoo SU(N)_{C'},
\eeq
where \GC\ is weakly coupled at the symmetry-breaking scale.
In this case, the surviving $SU(N)_{C'}$ group is weakly coupled, and we will
later take $N = 3$ and identify it with QCD.

To make these ideas specific, consider a theory with a gauge group
\eq
\label\gagroop
SU(N)_{TC} \times \big[ SU(K) \times SU(K) \big]_I
\times SU(N)_C,
\eeq
and fermion content
\eq\eqalign{
\psi_{Lj}, \psi_{Rj} & \sim (N, 1, 1, 1), \qquad j = 1, \ldots, F, \cr
\chi_L & \sim (N, K, 1, 1), \cr
\chi_R & \sim (1, K, 1, N), \cr
\xi_R & \sim (N, 1, K, 1), \cr
\xi_L & \sim (1, 1, K, N). \cr}
\eeq
For simplicity, we will impose a discrete symmetry to set the coupling
constants
of the two $SU(K)$ groups equal.
The first three group factors in eq.\ \gagroop\ are ``strong'' gauge groups
which will form fermion condensates, while $SU(N)_C$ is weakly coupled in the
energy regime in which we are interested.
Note that the fermion content of each gauge group consists of vector-like
fermions in the fundamental representation.
Thus, all gauge anomalies are cancelled in this theory.
More importantly, this means that as long as only one of the gauge couplings
is strong, we can analyze the symmetry breaking pattern by appealing to our
understanding of QCD-like theories.

The coupling strength of the gauge groups \GTC\ and \GI\ can be
characterized by the values of the gauge coupling constants $g_{TC}$ and
$g_I$ evaluated at some fixed scale $\mu_0$.
Alternatively, we can characterize the gauge couplings by mass scales
$\Lambda_{TC}$ and $\Lambda_I$ at which the respective gauge couplings become
strong enough to trigger chiral symmetry breaking.
There are several possibilities for the low-energy dynamics of this model,
depending on the relative magnitudes of these parameters.

$\bullet$ Suppose first that $g_I \gg g_{TC}$ at the scale
$\mu_0$, so that $\Lambda_I \gg \Lambda_{TC}$.
Then \GTC\ is weakly coupled at the scale $\Lambda_I$, and the
\GII\ interactions give rise to the condensates
\eq
\label\Icond
\avg{\mybar\chi_L \chi_R}, \avg{\mybar\xi_L \xi_R} \sim \Lambda^3_I.
\eeq
This condensate results in the chiral symmetry breaking pattern
\eq
\eqalign{
SU(N)_{\chi L} \times SU(N)_{\chi R} & \gotoo SU(N)_{\chi L + \chi R}, \cr
SU(N)_{\xi L} \times SU(N)_{\xi R} & \gotoo SU(N)_{\xi L + \xi R}, \cr}
\eeq
giving rise to $2(N^2 - 1)$ potential NGB's.
(Each of the $SU(K)$ factors acts like a copy of QCD.)
The condensate eq.\ \Icond\ results in the gauge symmetry breaking pattern
\eq
SU(N)_{TC} \times SU(N)_C \gotoo SU(N)_{C'}.
\eeq
$N^2 - 1$ potential NGB's are therefore eaten by the broken gauge bosons,
which acquire masses of order $g_{TC} \Lambda_I$.
The remaining $N^2 - 1$ potential NGB's transform in the adjoint representation
of the unbroken $SU(N)_{C'}$ gauge group, and therefore acquire masses of order
$g_C \Lambda_I$.

Through all of this, the $\psi$ fermions remain massless, and
$\avg{\mybar\psi\psi} = 0$.
The effective theory at scales $\mu \ll g_C \Lambda_I$ therefore
consists of the massless $\psi$ fermions subject to various contact
interactions suppressed by inverse powers of $\Lambda_I$.
This is qualitatively similar to the NJL model far into the unbroken phase.

$\bullet$ Now suppose that $g_{TC} \gg g_I$ at the scale $\mu_0$, so that
$\Lambda_{TC} \gg \Lambda_I$.
In this case, the \GTC\ gauge coupling becomes strong at a scale where the
\GI\ gauge coupling is weak.
This results in the formation of the condensate
\eq
\label\Ccond
\avg{\mybar\psi_L \psi_R + \mybar\chi_L \xi_R} \sim \Lambda_{TC}^3.
\eeq
(There are other possibilities for the alignment of the condensate, but a
calculation along the lines of ref.\ \align\ which treats the $SU(K)$ and
electroweak gauge groups perturbatively shows that this vacuum alignment is
chosen for sufficiently weak $SU(K)$ coupling.
This is consistent with the lore that condensates will align so as to preserve
the maximal gauge symmetry.)
This condensate results in the chiral symmetry-breaking pattern
\eq
\label\symbrek
SU(F + K)_L \times SU(F + K)_R \gotoo SU(F + K)_{L + R},
\eeq
giving rise to $(F + K)^2 - 1$ potential NGB's.
This gives rise to the gauge symmetry breaking pattern
\eq
\left[ SU(K) \times SU(K) \right]_I \gotoo SU(K)_{I'}.
\eeq
$K^2 - 1$ NGB's resulting from the condensate eq.\ \Ccond\ are therefore eaten
by the broken gauge bosons, which acquire masses $\sim g_I \Lambda_{TC}$.

In a basis where the first $F$ entries correspond to $\psi$ and the last $K$
entries correspond to $\chi_L$ and $\xi_R$, we can classify the potential
NGB's associated with the broken axial generators as follows:
\eqa
\label\physNGB
\pmatrix{X & 0 \cr 0 & 0 \cr}
& \qquad\hbox{$F^2 - 1$ physical NGB's,} \eol
\label\eatNGB
\pmatrix{0 & 0 \cr 0 & Y \cr}
& \qquad\hbox{$K^2 - 1$ eaten NGB's,} \eol
\label\extraNGB
\pmatrix{0 & E \cr E^\dagger & 0}
& \qquad\hbox{$2FK$ massive pseudo-NGB's,} \eol
\label\axion
\pmatrix{K \cdot 1 & 0 \cr 0 & -F \cdot 1 \cr}
& \qquad\hbox{$1$ physical ``axion.''} \eeol
\eeq
The fate of the NGB's of eqs.\ \physNGB\ and \eatNGB\ should be obvious.
The NGB's of eq.\ \extraNGB\ can be viewed as $2F$ NGB's transforming in the
fundamental representation of the unbroken \GII\ gauge group.
These NGB's therefore acquire masses of order $g_I \Lambda_{TC}$.
The ``axion'' of eq.\ \axion\ can be associated with the charge
\eq
\label\axq
\eqalign{
\psi_L \ : \ & +K, \qquad \psi_R \ : \ -K, \cr
\chi_L \ : \ & -F, \qquad \chi_R \ : \ -F, \cr
 \xi_L \ : \ & +F, \qquad  \xi_R \ : \ +F. \cr}
\eeq
The current associated with this charge has no
$SU(N)_{TC} \times \left[ SU(K) \times SU(K) \right]_I$ anomalies, but it
does have a $SU(N)_C$ anomaly.
Therefore, the axion will acquire a mass of order
$\Lambda_C^2 / \Lambda_{TC} \ll \Lambda_C$.

The $\chi_R$ and $\xi_L$ fermions remain massless at this stage, and
transform in the fundamental representation of the unbroken \GII\ gauge
group.
At the scale $\Lambda_{I'}$, the \GII\ gauge group becomes strong,
resulting in the formation of the condensate
\eq
\avg{\mybar\xi_L \chi_R} \sim \Lambda_{I'}^3.
\eeq
This results in the symmetry-breaking pattern
\eq
SU(N)_{\xi L} \times SU(N)_{\chi R} \gotoo SU(N)_{\xi L + \chi R},
\eeq
giving rise to $N^2 - 1$ potential NGB's.
These potential NGB's transform in the adjoint representation of the
unbroken $SU(N)_C$ gauge group, and therefore acquire masses of order
$g_C \Lambda_{I'}$.
Thus, $\avg{\mybar\psi\psi} \sim \Lambda_{TC}^3$ and the low-energy theory
consists entirely of the $F^2$ NGB's resulting from the symmetry breaking at
the scale $\Lambda_{TC}$.
This is qualitatively similar to the NJL model far into the broken phase.

$\bullet$ We now consider the nature of the transition between the two
limits just described.
If we {\it assume} that the order parameter $\avg{\mybar\psi_L\psi_R}$
is continuous across the transition (\ie\ that the transition is second
order), then we can tune the \GI\ coupling so that
$\avg{\mybar\psi_L \psi_R} \ll \Lambda^3$, where
$\Lambda \sim \Lambda_{TC} \sim \Lambda_I$.
In this case, both \GTC\ and \GI\ are becoming strongly coupled near the
scale $\Lambda$, but we can analyze this model using continuity arguments.

In the fine-tuned limit, it seems plausible that the $\psi$ fermions are
propagating degrees of freedom with mass $m \ll \Lambda$.
The reason is that in the unbroken phase, there are poles at zero momentum
transfer in Green's functions coming from an intermediate state consisting
of a single $\psi$ fermion.
If the transition is second order, we expect the position of these poles to be
continuous across the transition, and the $\psi$ fermions to be massive
propagating states.

Also, in the fine-tuned limit, we expect the decay constant $f$ of the
NGB's to be $f \sim m$.
To see this, define
\eq
\label\currcurr
\scr J^{\mu\nu}_{AB}(q) \equiv \myint d^4 x \; e^{-iq \cdot x}
\vev{T J^\mu_A(x) J^\nu_B(0)},
\eeq
where
\eq
J^\mu_A \equiv \mybar\psi i\gamma^\mu \gamma_5 T_A \psi
\eeq
are the spontaneously broken axial currents.
{}From Goldstone's theorem, we know that $\scr J^{\mu\nu}_{AB}(q)$
has a pole at $q^2 = 0$:
\eq
\label\ppole
\scr J^{\mu\nu}_{AB}(q) \gotoo ( q^\mu q^\nu - q^2 g^{\mu\nu} )
\frac{f_{AB}^2}{q^2}
\quad {\rm as} \quad q^2 \goto 0,
\eeq
where $f_{AB}^2$ is the matrix of NGB decay constants.
{}From the Jackiw--Johnson sum rule (eq.\ (54) below), we know that the NGB
decay constants vanish if the fermion self-energy vanishes.
Therefore, if we assume that the transition is second order, then we expect
the decay constants will be of order $m$ for $m \ll \Lambda$.
(This is made explicit in the approximations to the Jackiw--Johnson sum rule
which we will consider below.)

We now turn to the low-energy effective lagrangian for this model.
At sufficiently low energies $\mu \ll f$, the effective lagrangian describes
the interactions among the NGB's (and possibly the $\psi$ fermions if their
masses are smaller than $f$).
In this effective lagrangian, the $SU(N)_F$ symmetry is realized nonlinearly
on the NGB fields.
As a result, this effective lagrangian breaks down for processes involving
momentum transfers $p^2 \gg f^2$.
One signal for this fact is that an infinite number of operators are important
for such processes.
We therefore conclude that the theory {\it must} contain new particles and
interactions at scales $\mu \lsim f \ll \Lambda_{TC}$.
The simplest possibility is that the symmetry is realized linearly in the
effective lagrangian at scales $\mu \sim f$, implying that the theory contains
light scalars with the quantum numbers of the $\Phi$ field in eq.\ \efflin.
(It is difficult to imagine a reasonable alternative if one accepts
the assumption that the $\psi$ condensate can be fine-tuned to be small.)

In fact, we may conjecture that the effective lagrangian at scales below
$\Lambda_{TC}$ is in the same universality class as the NJL model.
To make this statement precise, we write the effective lagrangian as
\eq
\scr L\rsub{eff} = \scr L\rsub{NJL} + \delta\scr L
\eeq
where $\scr L\rsub{NJL}$ is the NJL lagrangian of eq.\ \NJLL\ and
$\delta\scr L$ contains all remaining terms.
We conjecture that the four-fermion coupling in $\scr L\rsub{NJL}$ is close to
its critical value, and that the coefficients of the operators in
$\delta\scr L$ are sufficiently small that the theory is in the same
universality class as the NJL model.
Similar assumptions have been discussed elsewhere for the case of
top-condensate
models \PZ\ and ``strong extended technicolor'' models \TCNJL.

\section{\SD\ Analysis}

\def\theprop#1{\frac{#1^2 \, \Sigma(#1^2)}
{#1^2 + \Sigma^2(#1^2)}}

We can flesh out our picture of the NJL limit by considering the
\SD\ equations for the theory we have described.
The full \SD\ equations are clearly intractable, and in order to
make progress we must truncate these equations and hope that what remains
captures the essential physics we wish to discuss.
We do not pretend that this analysis justifies the dynamical assumptions
made in the last section, but we will see that these assumptions are
incorporated in a simple and natural way in our analysis.
Also, we can use the truncated \SD\ equations to make crude quantitative
estimates for various quantities of interest.
This section can be skimmed by a reader interested mainly in the application of
the model discussed above to electroweak symmetry breaking.

\subsection{Fermion Self-Energy}

If we define
\eq
\Psi \equiv \pmatrix{\psi \cr \chi \cr \xi \cr},
\eeq
the full \SD\ equation for the $\Psi$ propagator is
\eq
\label\SDequ
iS^{-1}(p) = \sum_{r, r'} \myint \frac{d^4 k}{(2\pi)^4} \;
\big[ g_r \gamma^\mu T_{rA} \big] S(k)
\big[ \Gamma^\nu_{r'B}(k, p) T_{r'B} \big]
G_{\mu\nu}^{AB}(k - p),
\eeq
where $r, r'$ run over the fermion representations of the gauge group,
$\Gamma^\mu_{rA}$ is the gauge boson vertex function,
and $G_{\mu\nu}^{AB}$ is the gauge boson propagator.
We restrict the fermion propagator to have the block-diagonal form
\eq
\label\Sform
S = \pmatrix{S_\psi &\cr & S_{\chi\xi} \cr}.
\eeq
This is consistent with the full \SD\ equations, and is also the correct form
in both of the limits considered in the previous section.
The \SD\ equation for $S_\psi$ then involves only the exchange of
\GTC\ gauge bosons.
We will assume in what follows that the \GI\ coupling can be chosen so that
the mass $M$ of the \GTC\ gauge bosons has any desired value.

We will follow a venerable tradition and approximate the full gauge
boson vertex function by its tree-level value (with a running coupling
evaluated at an appropiate momentum scale), approximate the gauge
boson propagators by their asymptotic forms, and neglect fermion
wave-function renormalization.
(See \gaprev\ for more details.)
The resulting integral equation for the $\psi$ fermion self-energy is then
\eq
\label\gapequ
\Sigma(p^2) = \myint \frac{dk^2}{4\pi^2} \;
G(\max\{k^2, p^2, M^2\}) \;
\theprop{k},
\eeq
where $M$ is the mass of the $SU(N)_{TC}$ gauge bosons,
\eq
G(p^2) \equiv \frac{3 g^2(p^2) C_2}{p^2}
\eeq
can be viewed as the strength of the one-gauge-boson-exchange interaction,
and $C_2$ is the Casimir of the fermion representation.

For $p^2 \le M^2$, the \rhs\ of eq.\ \gapequ\ is independent of $p^2$, and we
have
\eq
\Sigma(p^2) = m \quad \hbox{for} \quad p^2 \le M^2.
\eeq
This situation is rather reminiscent of the NJL model.
In fact, we can write
\eq
\label\NJLgap
\eqalign{
m & = m \, \frac{G(M^2)}{4\pi^2} \int_0^{M^2} dk^2 \;
\frac{k^2}{k^2 + m^2} + \delta m \cr
& = m \, \frac{G(M^2)}{4\pi^2} \left[ M^2
- m^2 \ln\frac{M^2}{m^2} \right] + \delta m, \cr}
\eeq
where
\eq
\delta m \equiv \int_{M^2}^\infty \frac{dk^2}{4\pi^2} \; G(k^2) \; \theprop{k}
\eeq
depends only on the behavior of $\Sigma(p^2)$ with $p^2 > M^2$.
If we naively identify $G(M^2)$ with the NJL coupling $G$ and indentify
$M$ with the NJL cutoff $\Lambda$, then eq.\ \NJLgap\ is similar to the NJL
gap equation with the addition a ``counterterm'' $\delta m$.
To understand the relation between eq.\ \NJLgap\ and the NJL gap equation, we
note that the NJL coupling $G$ includes the effects of integrating out all
field models with $p^2 > \Lambda^2$, while $G(M^2)$ has no such
interpretation.
The role of the term $\delta m$ is to include the effects of the high-energy
gauge bosons.

For $p^2 > M^2$, eq.\ \gapequ\ is the same as the gap equation for
massless gauge bosons, which has been studied by many authors.
The integral equation can be converted to a differential equation:
\eq
\label\theDE
\Sigma''(p^2) - \frac{G''(p^2)}{G'(p^2)} \; \Sigma'(p^2)
- \frac{G'(p^2)}{4\pi^2} \Sigma(p^2) = 0,
\eeq
where the prime denotes differentiation with respect to $p^2$.
This equation has been linearized in $\Sigma$, which is completely justified
in the present case, since we are interested in the regime where
$p^2 \ge M^2 \gg \Sigma^2(p^2)$.
Using the known asymptotically-free behavior of $G(p^2)$ for large $p^2$, we
can write the ultraviolet boundary condition
\eq
\label\UVbdy
\Sigma(p^2) \gotoo \sigma^3 G(p^2)
\quad \hbox {as} \quad p^2 \goto \infty,
\eeq
for some constant $\sigma$.
Eq.\ \theDE\ has an infrared boundary condition
\eq
\label\IRbdy
\eqalign{
\Sigma'(M^2) & = m \, G'(M^2) \int_0^{M^2} \frac{dk^2}{4\pi^2} \;
\frac{k^2}{k^2 + m^2} \cr
& = m \, \frac{G'(M^2)}{4\pi^2}
\left[ M^2 - m^2 \ln\frac{M^2}{m^2} \right]. \cr}
\eeq
It is easily verified by direct substitution that eq.\ \theDE\ with these
boundary conditions is equivalent to eq.\ \gapequ.

In this formalism, we can explicitly see the fine-tuning necessary to
obtain $m \ll M$.
Because eq.\ \theDE\ is linear in $\Sigma$, once we impose the ultraviolet
boundary condition eq.\ \UVbdy, $\Sigma(p^2)$ is completely determined up to
an overall constant.
We can therefore write
\eq
\label\Sigmasoln
\Sigma(p^2) = m \; \frac{F(p^2)}{F(M^2)},
\eeq
where $F(p^2)$ is the solution to eq.\ \theDE\ satisfying the boundary
condition
\eq
F(p^2) \gotoo G(p^2) \quad\hbox{as}\quad p^2 \goto \infty.
\eeq
The infrared boundary condition can then be written
\eq
\label\IRtune
\frac{M^2 R(M^2)}{4\pi^2} \equiv
\frac{M^2 G'(M^2) F(M^2)}{4\pi^2 \, F'(M^2)} = 1 + O(m^2 / M^2).
\eeq
The left-hand side depends only on $M$ and is of order 1 for
$M \sim \Lambda_{TC}$.
We see that $M$ must be equal to a critical value $M_{\rm crit}$ to an
accuracy $O(m^2 / M^2)$ if we want $m \ll M$.
This in turn means that the \GI\ gauge coupling (evaluated at the scale
$\mu_0$) must be fine-tuned to an accuracy $O(m^2 / M^2)$.

To show that eq.\ \IRtune\ has a solution for some $M^2$, we note that the
quantity $R(p^2)$ defined above satisfies the differential equation
\eq
\label\RDE
R'(p^2) = G'(p^2) - R^2(p^2),
\eeq
with ultraviolet boundary condition
\eq
R(p^2) \gotoo G(p^2) \quad\hbox{as}\quad p^2 \goto \infty.
\eeq
Thus, the left-hand side of eq.\ \IRtune\ is small for $M \ll \Lambda_{TC}$.
Because $R'(p^2) < G'(p^2)$, we have $R(p^2) < G(p^2)$, and thus
\eq
M^2 R(M^2) > 3 g^2(M^2) C_2.
\eeq
Therefore, as long as $g^2(M^2)$ is sufficiently large,
the left-hand side of eq.\ \IRtune\ will be greater than unity and the
differential equation has a solution.
This is just the condition that the gauge coupling becomes sufficiently strong
to trigger chiral symmetry breaking.

Note also that $\Sigma(0)$ is continuous as a function of $M$ in this
formalism.

\subsection{Light Scalars}

We now address the question of the existence of light scalars in this model.
Our strategy is to compute $\mybar\psi$--$\psi$ scattering in the
color-singlet scalar and pseudo-scalar channels.
The light scalars will manifest themselves as poles in the $s$ channel.
We will compute the scattering amplitude in the ladder approximation shown in
fig.\ 1.

We will make several further simplifications.
Since the phase transition is second order in this formalism, it is sufficient
to establish the existence of the scalars in the {\it unbroken} phase where
$\Sigma \equiv 0$.
Also, we will compute the amplitude in the kinematic regime shown
in fig.\ 1.
The Mandlestam variables are then $s = 0$, $t = u = p^2$.
A light scalar with mass $m_s \ll M$ in the $s$ channel will result in the
behavior
\eq
T(p^2) \sim \frac 1{m_s^2} + \cdots.
\eeq
In this kinematic regime, only a single form-factor contributes to the
amplitude, and we can write the following simple integral equation for the
scattering amplitude $T(p^2)$:
\eq
\label\Tequ
T(p^2) = G(\max\{p^2, M^2\}) + \myint \frac{dk^2}{4\pi^2} \;
G(\max\{k^2, p^2, M^2\}) \, T(k^2).
\eeq
Since we are in the unbroken phase, this equation holds for all flavor channels
and for both the scalar and pseudoscalar channels.

For $p^2 \le M^2$, the right-hand side of eq.\ \Tequ\ is independent of $p^2$,
and we have
\eq
T(p^2) = T \quad\hbox{for}\quad p^2 \le M^2.
\eeq
Again, this situation is very similar to the NJL model, where the scattering
amplitude considered here is independent of $t$ and $u$.

For $p^2 > M^2$, the amplitude satisfies the differential equation
\eq
\label\TDE
T''(p^2) - \frac{G''(p^2)}{G'(p^2)} \; T'(p^2)
- \frac{G'(p^2)}{4\pi^2} \; T(p^2) = 0,
\eeq
where the prime again denoted differentiation with respect to $p^2$.
This is the same differential equation satisfied by $\Sigma$ above, but with
different boundary conditions and with the crucial difference that it holds
for arbitrarily large $T$.
Eq.\ \TDE\ has the ultraviolet boundary condition
\eq
T(p^2) \gotoo cG(p^2) \quad\hbox{as}\quad p^2 \goto \infty,
\eeq
where $c$ is a dimensionless constant, and an infrared boundary condition
\eq
T'(M^2) = G'(M^2) \left[ 1 + \frac{T(M^2) M^2}{4\pi^2} \right].
\eeq
Once we impose the ultraviolet boundary condition, we can write the
solution as
\eq
T(p^2) = T \; \frac{F(p^2)}{F(M^2)},
\eeq
where $F(p^2)$ is the function introduced in eq.\ \Sigmasoln\ above.
The infrared boundary condition can then be written
\eqa
\frac 1T & = \frac{F'(M^2)}{G'(M^2) F(M^2)} - \frac{M^2}{4\pi^2} \eolnn
\label\theT
& = \frac 1{R(M^2)} - \frac{M^2}{4\pi^2} \eeol
\eeq
where $R(p^2)$ is the function defined in eq.\ \IRtune.
Eq.\ \theT\ shows explicitly that a solution exists.
Of more interest is the fact that eq.\ \theT\ shows that as we approach the
fine-tuning condition eq.\ \IRtune, $T$ becomes singular.\footnote{$^\dagger$}
{In an exact treatment, we certainly expect the same critical value of $M$ for
the chiral symmetry-breaking transition and the appearance of light scalars,
but {\it \'a priori} we have no right to expect the critical values of $M$
to concide exactly in our approximation, since it is not clear that the
gap equation eq.\ \gapequ\ and the partial summation we have performed to
compute $T$ are part of a single consistent approximation.}
This shows that there are light scalars in the theory near the critical
point in the ladder approximation.

\subsection{Decay Constants}

We now discuss the decay constants which result from the $\psi$ fermion
condensation.
They are given by the Jackiw--Johnson sum rule \JJ
\eq
\label\JJsum
f_{AB}^2 = \myint \frac{d^4 k}{(2\pi)^4} \;
\tr\left[ \gamma_\mu \gamma_5 (Z_J T_A) S(k)
\twi\Gamma^\mu_B(k, k) S(k) \right] + \delta f^2,
\eeq
where $\twi\Gamma^\mu_A$ is the current vertex function with the NGB pole
removed, $S(k)$ is the propagator of the fermions in the current,
$Z_J$ is the renormalization constant of the current operator,
and $\delta f^2$ is a counterterm.

If we make the same approximations made in deriving eq.\ \gapequ\ we
obtain $f^2_{AB} = f^2 \delta_{AB}$, with
\eqa
f^2 & \simeq \frac N{16\pi^2} \myint dk^2 \;
\frac{\Sigma^2(k^2) \left[ 2k^2 + \Sigma^2(k^2) \right]}
{\left[ k^2 + \Sigma^2(k^2) \right]^2} \eol
\label\thef
& = \frac N{16\pi^2} \; \left[ m^2 \, \ln\frac{M^2}{m^2}
+ \int_{M^2}^\infty \; dk^2 \; \frac{\Sigma^2(k^2)}{k^2} \right]. \eeol
\eeq
The integral in eq.\ \thef\ is of order $m^2$, so the first term dominates
for $m \ll M$.
If we approximate $\Sigma(p^2)$ for $p^2 > M^2$ by its asymptotic form
eq.\ \UVbdy, we obtain
\eq
\label\theff
f^2 \simeq \frac{N m^2}{16\pi^2} \;
\left[ \ln\frac{M^2}{m^2} + \frac 12 \right].
\eeq

To summarize, we see that the gap-equation analysis incorporates many of the
features we expected on the basis of qualitative arguments of section 3:
the transition is second order, the fine-tuning is manifest, the decay
constants are small compared to the compositeness scale, and the light scalars
are present.

\section{Application to Electroweak Symmetry Breaking}

\subsection{``Top-Mode'' Model}

In order to apply the model described above to electroweak symmetry breaking,
we take $N = 3$ and identify $SU(3)_C$ with ordinary color.
We also take $F = 2$ and identify
\eq
\psi = \pmatrix{t \cr b \cr}.
\eeq
The low energy theory contains two Higgs doublets; in the notation of
eq.\ \efflin, we have
\eq
\Phi = \pmatrix{H_t^0 & H_b^+ \cr H_t^- & H_b^0 \cr},
\eeq
where $H_t$ and $H_b$ are Higgs doublet fields.

This theory may seem to be a phenomenological disaster, since it appears
that custodial symmetry of the technicolor interactions implies $m_t = m_b$,
and the theory contains an axion (see eq.\ \axq) with a decay constant near
the weak scale.
Such axions are ruled out by a combination of laboratory experiments and
astrophysical considerations \axionrev.
However, the model actually does not suffer from these problems, as we will
describe below.

In the standard model, we know that the custodial $SU(2)$ symmetry is broken
by the weak hypercharge.
In fact, the hypercharge assignments are
\eq
Y(\psi_L) = \sfrac 13, \quad
Y(t_R) = \sfrac 43, \quad
Y(b_R) = -\sfrac 23.
\eeq
Thus, $U(1)_Y$ gauge boson exchange will mediate an {\it attractive} force
between $\psi_L$ and $\mybar t_R$ and a {\it repulsive} force between
$\psi_L$ and $\mybar b_R$.
Although this force is weak, when the theory is tuned near the critical point,
the contribution from $U(1)_Y$ gauge boson exchange can result in
$\avg{\mybar bb} = 0$ while $\avg{\mybar tt} \ne 0$.
The phenomenon of the amplification of small perturbations near a critical
point has been discussed in the context of technicolor theories \crit\ and
in top-condensate models \ttcrit, where it has been dubbed ``critical
instability.''

To see how this phenomenon emerges from the gap equation, write
\eqa
Y & = Y_L P_L + Y_R P_R \eol
& = \sfrac 13 1_2 P_L + \left( \sfrac 13 1_2 + \tau_3 \right) P_R,
\eeq
where $P_{L, R}$ are the left- and right-handed helicity projection operators.
The $\psi$ self-energy is a diagonal $2 \times 2$ matrix
\eq
\Sigma = \pmatrix{\Sigma_t & 0 \cr 0 & \Sigma_b \cr}.
\eeq
The gap equation has the same form as eq.\ \gapequ\ with the replacement
\eq
G(p^2) = \frac 3{4\pi^2 p^2} \left[ g_{TC}^2(p^2) C_2
+ g_Y^2(p^2) Y_L Y_R \right].
\eeq
In the approximations we are making, the Schwinger--Dyson equations for the
fermion masses split into separate equations for the top- and bottom-quark
self-energies.
Suppose now that the theory is tuned so that $m_t \ll M$.
Then
\eq
\label\thecrit
\eqalign{
G_t(M^2) & = G_{\rm crit} + O(m_t^2 / M^2), \cr
G_b(M^2) & = G_{\rm crit} - \frac{2 g_Y^2(M^2)}{M^2}
+ O(m_t^2 / M^2). \cr}
\eeq
Thus, $\Sigma_b = 0$ as long as
\eq
g_Y \roughly> \frac{m_t}M.
\eeq
(Note that we can consider the possibility that the scale $M$ is chosen so
that the hierarchy $m_t / m_b$ emerges entirely as a result of $U(1)_Y$
custodial symmetry breaking.
However, this requires $M \sim 1 \TeV$, and results in a top mass
$m_t \sim 500 \GeV$, as we will see below.)

We now consider the low-energy theory for the case where $\Sigma_b = 0$.
Taking into account the breaking of $SU(2)_{\psi R}$ due to hypercharge,
the symmetry breaking pattern can be written
\eq
SU(2)_W \times U(1)_Y \times U(1)_{bR}
\gotoo U(1)_{EM} \times U(1)_{bR},
\eeq
resulting in 3 potential NGB's which are eaten by the massive electroweak
gauge bosons.
The effective theory below the scale $\Lambda_{TC}$ contains two Higgs
doublets with
\eq
\avg{H_t} = v, \qquad
\avg{H_b} = 0.
\eeq
$H_b$ is an unbroken doublet with mass $\sim g_Y M$, which can be far above
the electroweak scale.
In this model as discussed so far, all fermions other than the top quark are
exactly massless.

The axion of eq.\ \axion\ is not present, but this is only because there is an
unbroken axial $U(1)$ symmetry in the low-energy theory as a result of the
massless $b$ quark.
Clearly, this symmetry, as well as the flavor symmetries of the remaining
quarks and leptons, must be broken somehow if the theory is to account for the
observed fermion masses and mixings.
This is exactly the situation in technicolor theories, where the technifermion
condensate does not give mass to ordinary quarks and leptons in the absence of
additional ``extended technicolor'' interactions which connect the
technifermions with the quarks and leptons.

Thus, as in technicolor, we assume that the effective theory at the scale
$\Lambda_{TC}$ contains four-fermion ``extended technicolor'' (ETC)
interactions
of the form
\eq
\label\ffermi
\frac 1{M_E^2} \; (\mybar\psi \psi) (\mybar ff)
\eeq
where the $\psi$'s represent third-generation quark fields and the $f$'s
represent quark or lepton fields which are $SU(N)_{TC}$ singlets.
We will not be concerned here with the dynamics which gives rise to these
operators, or with exhibiting a completely realistic theory.
Our aim is simply to show that there is in principle no obstacle to making
the theory realistic.

Four-fermion interactions of the form of eq.\ \ffermi\ can give rise to masses
for all the quarks and leptons, but they necessarily preserve a $U(1)$
symmetry which counts the number of quarks from the first two
generations.\footnote{$^\dagger$}
{I am indebted to R.\ Sundrum for pointing this out to me.}
If this symmetry were exact, elements of the Cabbibo--Kobayashi--Maskawa matrix
involving the third generation would vanish identically, and so this $U(1)$
symmetry must be broken somehow.
This $U(1)$ symmetry can be broken by dimension-7 interactions such as
\eq
\frac 1{M_E^3} \; (\mybar Q_L i\Sla D \psi_L)
(\mybar\chi_L \chi_R),
\eeq
where $Q_L$ is a left-handed quark field from the first two generations.
Alternatively, we can imagine that {\it all} quarks carry $SU(3)_{TC}$ rather
than $SU(3)_C$, and the fact that the top quark is the only fermion which
condenses is explained by critical instability, due for example to additional
higher-dimension operators.
In this case, all masses and mixings can be obtained from operators of the form
of eq.\ \ffermi.
Thus, there seems to be no obstacle in principle to incorporating light
fermions into the theory.

The scale $M_E > \Lambda_{TC}$ is associated with new interactions, such as
massive gauge boson exchange.
It may seem that much of the motivation for writing a renormalizable model
is lost once we introduce these non-renormalizable terms.
However, unlike the four-fermi coupling which drives the symmetry breaking in
the NJL model, these four-fermi interactions are perturbatively weak, and
no exotic dynamics is required to generate them.
The present model can be viewed as a valid effective theory for scales up to
$\sim M_E$, which can be well above the compositeness scale, while the NJL
model cannot be extended above scales $\Lambda \sim 1 / \sqrt{G}$.

Through the standard ETC mechanism, the four-fermi interactions of
eq.\ \ffermi\ will give rise to fermion masses of order
\eq
\label\fermass
m_f \sim \frac 1{M_E^2} \; \int^M \frac{d^4 k}{(2\pi)^4} \;
\tr\frac 1{\sla k} \; m_t \; \frac 1{\sla k}
\sim \frac{M^2 m_t}{4\pi^2 M_E^2}.
\eeq
This can be viewed as due to the presence of a condensate
\eq
\avg{\mybar\psi\psi} \sim \frac{M^2 m}{4\pi^2}.
\eeq
Alternatively, using the NJL model as a guide, we expect that the ETC
interactions eq.\ \ffermi\ will give rise to Yukawa couplings between the
ordinary fermions and the composite Higgs field of order
\eq
y \sim \frac{M^2}{4\pi^2 M_E^2} \; \frac{m_t}{v}
\eeq
This gives rise to the fermion masses eq.\ \fermass\ when the Higgs field
acquires a vacuum expectation value.

We now consider the case where $M$ is far above the weak scale.
In this case, the effective lagrangian at low scales is indistinguishable from
the standard model, and we expect the renormalization group analysis of
ref.\ \BHL\ to be valid.
This scenario necessitates a severe fine-tuning of the order
$m_t^2 / \Lambda_{TC}^2$, and results in a top-quark mass of $230 \GeV$
for $\Lambda_{TC} \simeq 10^{15} \GeV$ ($200 \GeV$ for
$M_{TC} \simeq 10^{19} \GeV$).
Such values for the top quark mass are disfavored by a global analysis of the
radiative effects in the the standard model, which gives $m_t < 201 \GeV$
at the $95\%$ confindence level \PDG.

Note that if $M$ is far above the weak scale, eq.\ \fermass\ shows that
realistic fermion masses can easily be generated with $M_E$ sufficiently
large to suppress flavor-changing neutral currents which can arise from
four-fermion operators of the form
\eq
\frac 1{M_E^2} \; (\mybar ff) (\mybar ff).
\eeq
Of course, this feature is obtained only at the expense of the fine-tuning of
the $SU(N)_{TC}$ breaking scale.

If $M$ is close to the weak scale, the leading-log effects which form the
basis of the predictions of ref.\ \BHL\ are not expected to be important.
Using eq.\ \theff\ and noting that $v = f / \sqrt{2} = 246 \GeV$, we find \eg
\eq
m_t \simeq 440 \GeV \quad\hbox{for}\quad M \simeq 10 \TeV.
\eeq
While the approximations used to obtain this result are rather crude, it is
clear that this is well outside the limits on the top mass coming from
$\rho$-parameter constraints.
It is of course possible that other sources of custodial symmetry
breaking (for example from higher-dimensional operators from the ETC sector)
can cancel the effects of the top quark on the $\rho$ parameter, although there
is no good reason to expect this to happen.

Since the ``top-mode'' model described above seems to have phenomenological
troubles for all reasonable choices of parameters, we now turn to models
in which other fermions are responsible for the electroweak symmetry breaking.

\subsection{Technicolor-like Models}

In this subsection, we consider models in which the $\psi$ fermions are new
strongly-interacting fermions.
In this case, the direct connection between the top-quark mass and the
electroweak scale is lost, but raising the compositeness scale may make these
theories more phenomenologically attractive than technicolor theories (at the
expense of fine-tuning).
We will argue that because of the critical instability mechanism discussed
above, we cannot have the electroweak symmetry broken by new fermions without
additional fine-tuning.

There are several strong constraints on the new fermions, which we will
generically denote by ``$T$.''
We assume that the new fermions occur in weak doublets so that the condensate
of these fermions is an $I_3 = \frac 12$ order parameter.
The masses of the $T$ fermions are constrained by demanding that they give
rise to the correct value of the electroweak scale $v$.
Using the approximation eq.\ \theff\ to the Jackiw--Johnson sum rule, we obtain
\eqa
\sum_j m_j^2 & \simeq \frac{8\pi^2 v^2}3 \;
\left( \ln\frac{M^2}{\mybar m^2} + \frac 12 \right)^{-1} \eol
& \sim (400 \GeV)^2 \quad\hbox{for}\quad M \sim 10 \TeV, \eeol
\eeq
where the sum runs over all $T$ fermions and $\mybar m$ is their average mass.
(This formula will have large corrections if $\ln (M^2 / \mybar m^2) \gg 1$.)
This shows that in any such model, we expect new heavy weak-doublet fermions
which will be accessible in future experiments.

In order that they do not contribute significantly to the
$\rho$-parameter, $T$ fermions in the same weak doublet must be nearly
degenerate in mass.
In particular, we must avoid large mass splittings induced by $U(1)_Y$
couplings through critical instability, as discussed above.
Of course, we can always add additional interactions to cancel this effect,
but the strength of these interactions must be fine-tuned, and we want to
see whether fine-tuning can be avoided.

To avoid the critical instability mechanism described in the last section,
we must take the weak doublet $T$ fermions to carry zero $U(1)_Y$ charge.
The \GW\ singlet $T$ fermions must carry $U(1)_Y$ charge so that the
condensate $\avg{T_L T_R}$ preserves $U(1)_{EM}$.
Thus, the only possibility is to take the extra fermions to transform under
$SU(3)_{TC} \times SU(2)_W \times U(1)_Y$ as
\eq
\eqalign{
T_L & \sim (3, 2; 0), \cr
U_R & \sim (3, 1; 1), \cr
D_R & \sim (3, 1; -1). \cr}
\eeq
When $SU(3)_{TC}$ becomes strong, they give rise to a condensate
\eq
\avg{\mybar U_L U_R}, \avg{\mybar D_L D_R} \ne 0,
\eeq
where we have defined
\eq
T_{L,R} \equiv \pmatrix{U \cr D \cr}_{L, R}.
\eeq
The theory has an approximate {\it discrete} custodial symmetry
\eq
\eqalign{
T_{L,R} & \mapsto \pmatrix{0 & 1 \cr 1 & 0 \cr} T_{L,R}
\equiv X T_{L,R}, \cr
B_\mu & \mapsto -B_\mu, \cr
W_\mu & \mapsto X W_\mu X, \cr}
\eeq
where $B_\mu$ and $W_\mu$ are the $U(1)_Y$ and $SU(2)_W$ gauge fields,
respectively.
If the $T$ fields were the only fermions in the theory, this custodial
symmetry would be exact, and we would have $m_U = m_D$.
In the full theory, this symmetry is violated by the $U(1)_Y$ gauge couplings
to the quarks and leptons and by ETC operators.
In particular, there must be strong ETC operators to give rise to the large
top mass, so that there will in general be custodial-symmetry violating
four-fermi couplings among the $T$ fermions with strength
$1 / M_E^2 \sim 1 / M^2$, which will give rise to a large splitting
$|m_U - m_D|$.
(Note that we cannot explain the large top-quark mass by taking the
third-generation quarks to transform under \GTC, since the effect of $U(1)_Y$
would give $m_t \gg m_T$.)

We have argued that additional fine-tuning is required in ``non-minimal''
models
incorporating NJL-like dynamics to break electroweak symmetry in order to avoid
an unacceptably large value for the $\rho$-parameter.
If we accept this additional fine-tuning, however, there seems to be no
obstacle to making such theories fully realistic.

\section{Conclusions}

We have considered a renormalizable model which we argued can break the
electroweak symmetry via NJL-like dynamics.
The crucial dynamical assumption is that the chiral symmetry-breaking
transition in the model is second order.
The fine-tuning needed to obtain a large hierarchy between the weak scale and
the compositeness scale is mainfested in the fine-tuning of a gauge coupling
constant.
We have seen that the minimal ``top-mode'' standard model can be obtained from
such a model, as well as models in which new fermions are responsible for
electroweak symmetry breaking.
In the latter case, we have argued that additional fine-tuning is required to
avoid unacceptably large corrections to the electroweak $\rho$-parameter.

\section{Acknowledgements}

I would like to thank R.\ Rattazzi and especially R.\ Sundrum for discussions
on the topic of this paper.
This work was supported by the Director, Office of Energy
Research, Office of High Energy and Nuclear Physics, Division of High
Energy Physics of the U.S. Department of Energy under Contract
DE-AC03-76SF00098.


\listrefs
\vfill
\eject
\centerline{\bf Figure Captions}
\vskip .4in
\noindent
Fig.\ 1:\ Ladder approximation to the $\psi$--$\mybar\psi$ scattering
amplitude.

\bye